\pdfoutput=1
\documentclass[10pt,english]{scrartcl}
\usepackage{helvet}

\usepackage[T1]{fontenc}
\usepackage[latin9]{inputenc}
\usepackage[letterpaper]{geometry}
\usepackage{babel}
\usepackage{float}
\usepackage{units}
\usepackage{amsbsy}
\usepackage{graphicx}
\usepackage{esint}
\usepackage[unicode=true,pdfusetitle,
 bookmarks=false,
 breaklinks=false,pdfborder={0 0 0},backref=false,colorlinks=false]
 {hyperref}

\makeatletter

\providecommand{\tabularnewline}{\\}

\renewcommand{\fnum@figure}{Fig.~\thefigure}

\makeatother

\begin{document}

\title{Conditions for the invertibility of dual energy data}

\author{Robert E. Alvarez\thanks{\protect\href{http://www.aprendtech.com}{aprendtech.com}, ralvarez@aprendtech.com}}
\maketitle
\begin{abstract}
The Alvarez-Macovski method {[}Alvarez, R. E and Macovski, A., \textquotedbl{}Energy-selective
reconstructions in X-ray computerized tomography\textquotedbl{}, Phys.
Med. Biol. (1976), 733--44{]} requires the inversion of the transformation
from the line integrals of the basis set coefficients to measurements
with multiple x-ray spectra. Analytical formulas for invertibility
of the transformation from two measurements to two line integrals
are derived. It is found that non-invertible systems have near zero
Jacobian determinants on a nearly straight line in the line integrals
plane. Formulas are derived for the points where the line crosses
the axes, thus determining the line. Additional formulas are derived
for the values of the terms of the Jacobian determinant at the endpoints
of the line of non-invertibility. The formulas are applied to a set
of spectra including one suggested by Levine that is not invertible
as well as similar spectra that are invertible and voltage switched
x-ray tube spectra that are also invertible. An iterative inverse
transformation algorithm exhibits large errors with non-invertible
spectra.

\end{abstract}

\section{Introduction}

In the Alvarez-Macovski\cite{Alvarez1976} method for extracting x-ray
spectral data, the attenuation coefficient is expanded as a linear
combination of functions of energy multiplied by basis set coefficients
that depend on the composition of the material at points within the
object. The line integrals of the basis set coefficients are then
computed by inverting the transformation from the measurements with
multiple spectra. Summarizing the measurements by the vector $\mathbf{L}$
and the line integrals of the basis set coefficients as the vector
$\mathbf{A}$, we need to show that the transformation $\mathbf{L(A)}$
is invertible. This question is examined in this paper. In particular,
the transformation for a noise-free system using two measurement spectra
with a two function basis set, \emph{i.e.} a dual energy system, is
studied.

Invertibility is of more than mathematical interest since spectra
that are near non-invertible can lead to a large multiplication of
noise in the output images. In the past, dual energy systems used
measurements with two different x-ray tube voltages or a fixed tube
voltage and filtering with two materials both with energy integrating
detectors. Experience showed that these were invertible. However,
the introduction of photon counting detectors into clinical systems\cite{taguchi2013vision}
leads to potential invertibility problems due to the imperfections
in these detectors\cite{Overdick2008} such as pileup, K-radiation
escape, charge sharing, and others. My recent paper\cite{alvarez2017noninvertibility}
showed that a large pileup factor can lead to non-invertibility for
a three dimension A-vector with three bin pulse height analysis and
that this non-invertibility leads to large increases in noise.

I discussed invertibility in other papers. In early work\cite{Alvarez1982,Lehmann1986},
I showed that a general, two dimensional inversion theorem\cite{Fulks1978}
could be applied to the dual energy transformation. The theorem requires
that the transformation be invertible on a closed contour in the first
quadrant of the A-vector plane and have a non-zero Jacobian determinant
inside the contour. I showed that the transformation is invertible
on the contour if the two spectra have different maximum energies.
The proof was for ordinary energy integrating detectors as well as
photon counting detectors without pileup. Later, I showed that the
transformation is invertible on the contour for photon counting detectors
with pileup\cite{alvarez2017DEInvertibilityProof}. 

Recently, Levine\cite{levine2017nonuniqueness} described a different
approach to determining invertibility. He studied a system with energy
spectra with only three discrete energies, no scatter and a perfect
detector. He found that, with appropriately chosen spectral values,
the transformation is not one-to-one and is therefore not invertible.
The spectra had the same maximum energy so my previous results were
not applicable. 

My papers did not address the proof that the Jacobian determinant
is non-zero within the contour. In this paper, I derive analytical
formulas are derived to test for the non-zero condition for a set
of measurement spectra and detector. The formulas are applied to several
systems including (1) the set proposed by Levine\cite{levine2017nonuniqueness}
that are not invertible as well as (2) similar spectra that are invertible
and (3) invertible voltage switched x-ray tube spectra that have long
been used in dual energy imaging. The results are verified by testing
the errors with an iterative algorithm that inverts the transformation.

\section{Methods}

\subsection{The transformation\label{sub:Alv-Mac-method}}

This section reviews the Alvarez-Macovski method\cite{Alvarez1976},
introduces notation, and specifies the transformation that is required
to be inverted for the method. 

For biological materials, we can approximate the x-ray attenuation
coefficient $\mu(\mathbf{r},E)$ with a two function basis set\cite{alvarez2013dimensionality}
\begin{equation}
\mu(\mathbf{r},E)=a_{1}(\mathbf{r})f_{1}(E)+a_{2}(\mathbf{r})f_{2}(E).\label{eq:2-func-decomp}
\end{equation}

\noindent{}In this equation, the coefficients $a_{i}(\mathbf{r})$
are functions only of the material composition at a position $\mathbf{r}$
within the object and the functions $f_{i}(E)$ depend only on the
x-ray energy $E$. If there is a high atomic number contrast agent,
we can extend the basis set to higher dimensions. 

Neglecting scatter, the expected value of the number of transmitted
photons $\lambda_{k}$ with an effective measurement spectrum $S_{k}(E)$
is 
\begin{equation}
\lambda_{k}=\int S_{k}(E)e^{-\int_{\mathcal{L}}\mu\left(\mathbf{r},E\right)d\mathbf{r}}dE\label{eq:Ik-integral}
\end{equation}

\noindent{}where the line integral in the exponent is on a line $\mathcal{L}$
from the x-ray source to the detector. The effective measurement spectra
include the effects of the source spectrum and the detector response.
Using the attenuation coefficient decomposition, Eq. \ref{eq:2-func-decomp},
the line integral is 
\begin{equation}
\int_{\mathcal{L}}\mu\left(\mathbf{r},E\right)d\mathbf{r}=A_{1}f_{1}(E)+A_{2}f_{2}(E).\label{eq:L(E)-A1-f1-A2-f2}
\end{equation}

\noindent{}where $A_{i}=\int_{\mathcal{L}}a_{i}\left(\mathbf{r}\right)d\mathbf{r},\ i=1\ldots2$
are the line integrals of the basis set coefficients. If the $A_{i}$
are summarized as the components of the A-vector, $\mathbf{A}$, and
the basis functions at energy $E$ by the vector $\mathbf{f(E)}$,
the expected values in Eq. \ref{eq:Ik-integral} are
\begin{equation}
\lambda_{k}\left(\mathbf{A}\right)=\int S_{k}(E)e^{-\mathbf{A\cdot f(E)}}dE.\label{eq:lambda-k}
\end{equation}
This paper assumes noise free data so the measurements are the expected
values. These measurements are summarized by a vector, $\mathbf{N}$,
whose components are the expected value of the photon counts with
each effective spectrum. 

Since the body transmission is exponential in $\mathbf{A}$, we can
approximately linearize the measurements by taking logarithms. The
results is the log measurement vector
\begin{equation}
\mathbf{L(A)}=-\log(\mathbf{\frac{N(A)}{\mathbf{\boldsymbol{N}}_{0}}})\label{eq:L-definition}
\end{equation}
\noindent{}where $\mathbf{\boldsymbol{N}_{0}}$ is the expected value
of the measurements with no object in the beam and the division means
that corresponding members of the vectors are divided. 

Eq. \ref{eq:L-definition} defines a transformation from the log measurements
$\mathbf{L(A)}$ to the A-vector, $\mathbf{A}$.

\subsection{The Invertibility Theorem}

The following theorem (Fulks\cite{Fulks1978} page 284) gives conditions
for any two dimension transformation to be invertible:
\begin{quotation}
Let F be a continuously differentiable mapping defined on an open
region D in E2, with range R in E2, and let its Jacobian be never
zero in D. Suppose further that C is a simple closed curve that, together
with its interior (recall the Jordan curve theorem), lies in D, and
that $\mathbf{F}$ is one-to-one on C. Then the image $\Gamma$ of
C is a simple closed curve that, together with its interior, lies
in R. Furthermore, $\mathbf{F}$ is one-to-one on the closed region
consisting of C and its interior, so that the inverse transformation
can be defined on the closed region consisting of $\Gamma$ and its
interior.
\end{quotation}
\noindent{}In Fulks' notation, $E2$ is a two dimensional Euclidean
space, so this theorem is applicable to the transformation $\mathbf{L(A)}$
for the two-measurement, two-basis function case.

I paraphrase the theorem as
\begin{quotation}
If the Jacobian of a continuously differentiable two dimensional mapping
is nonzero throughout an open region D and if the mapping is one to
one on a simple closed curve C which lies in D, then the mapping is
one to one and invertible on C and its interior. 
\end{quotation}
\noindent{}Fulks uses the term ``Jacobian'' to refer to the determinant
of the Jacobian matrix
\[
J=\det\left[\frac{\partial L_{i}}{\partial A_{j}}\right].
\]

I will assume that the converse of the theorem is also true. That
is, if the Jacobian determinant is zero in a region, then the transformation
is not invertible in that region. This is reasonable given the one-dimension
case where a zero derivative implies the transformation is not one-to-one
in a region around the zero value and is therefore not invertible.

\begin{figure}
\includegraphics[scale=0.6]{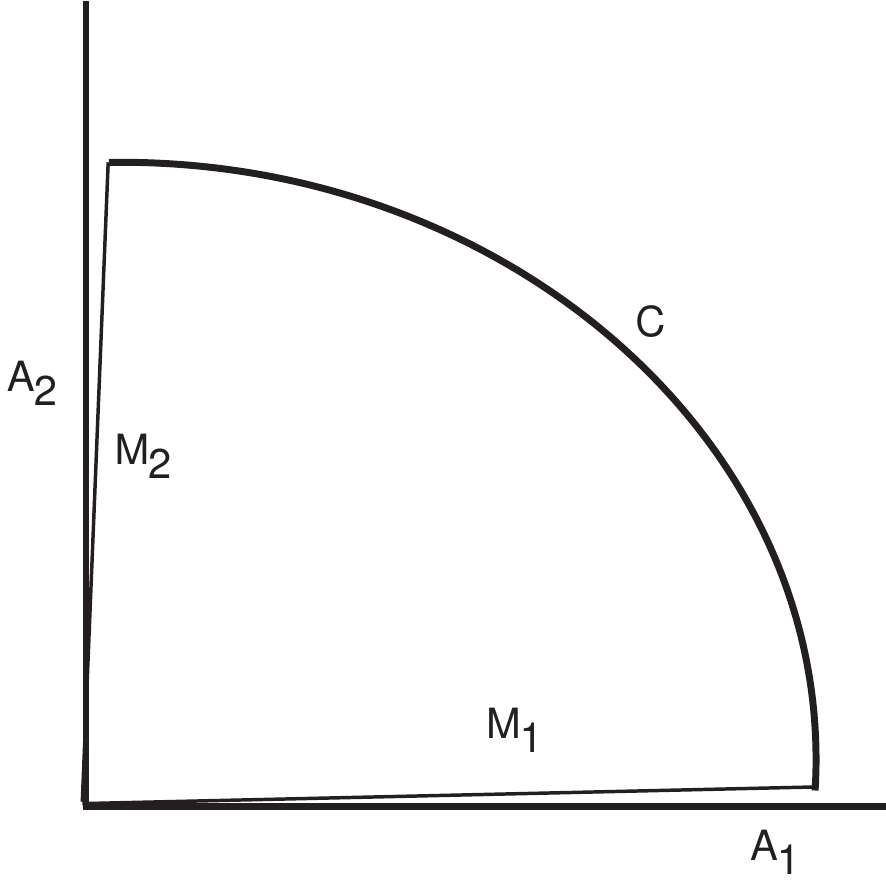}\centering{}

\protect\caption{Closed contour used in proof of invertibility. The contour consists
of the two axes, drawn slightly offset here to make them visible,
and a quadrant of a circle centered at the origin joining the endpoints
of the line segments along the axes.\label{fig:Closed-contour-4InvertProof-1}}
\end{figure}

I discussed the application of this theorem to $\mathbf{L(A)}$ in
previous papers\cite{Alvarez1982,Lehmann1986,alvarez2017DEInvertibilityProof}.
The proofs in these papers used the contour $C$, shown in Fig. \ref{fig:Closed-contour-4InvertProof-1},
which consists of the $A_{1}$ and $A_{2}$ axes and a section of
a circle centered on the origin and joining the end points of the
segments on the axes. In the previous papers, I proved that $\mathbf{L(A)}$
is invertible on the $A_{1}$ and $A_{2}$ axes under very general
conditions since objects whose A-vectors are on these axes are composed
of only one material. 

The previous invertibility proofs on the circular portion of $C$
are not applicable to some of the spectra in this paper. The proof
assumes that the maximum energies in the spectra are different but
the spectra discussed below have the same three energies but with
different weights. The previous argument was that, due to beam hardening,
as the radius of the contour becomes larger the transmitted spectrum
approaches a peak at the maximum energy. If the maximum energies of
the measurement spectra are different, then the transformation will
be invertible on the circle. However, if the maximum energies are
the same, the transformation becomes non-invertible for large object
thicknesses by the No-Linearize theorem (see Section 6.4 of my dissertation\cite{Alvarez}
and Chapter 44 of my book\cite{AlvarezBlogBook}). However, the transformation
may be invertible for moderate thicknesses if we can guarantee that
the circular portion of the contour with a smaller radius is in a
region where the Jacobian is non-zero. 

\subsection{Interpretation of the M Matrix and its derivative}

For our case, the Jacobian matrix is the matrix $\mathbf{M}$ with
coefficients 
\begin{equation}
M_{ij}=\frac{\partial L_{i}}{\partial A_{j}}.\label{eq:Mij}
\end{equation}
Since $L_{i}=-\log\left(\frac{\lambda_{i}}{\lambda_{i,0}}\right)$,
we can rewrite the elements of $\mathbf{M}$ as
\begin{equation}
M_{ij}=-\frac{1}{\lambda_{i}}\frac{\partial\lambda_{i}}{\partial A_{j}}.\label{eq:Mij-1}
\end{equation}
Substituting the expected value of measurements Eq. \ref{eq:lambda-k}
in this equation
\begin{equation}
M_{ij}=\frac{\int f_{j}(E)S_{i}(E)e^{-\mathbf{A\cdot f(E)}}dE}{\int S_{i}(E)e^{-\mathbf{A\cdot f(E)}}dE}.\label{eq:Mij-integrals}
\end{equation}

\noindent{}Defining the normalized spectrum
\[
\hat{s_{i}}(E)=\frac{S_{i}(E)e^{-\mathbf{A\cdot f(E)}}}{\int S_{i}(E)e^{-\mathbf{A\cdot f(E)}}dE}
\]

\begin{equation}
M_{ij}=\int f_{j}(E)\hat{s}_{i}(E)dE=\left\langle f_{j}(E)\right\rangle _{i}.\label{eq:M-ij-fj-eff}
\end{equation}

\noindent{}That is, $M_{ij}$ is the effective value of the $f_{j}(E)$
basis function in the transmitted measurement spectrum $S_{i}(E)e^{-\mathbf{A\cdot f(E)}}$. 

The derivative of $M_{ij}$ with respect to each of the A-vector components
can be computed by differentiating Eq. \ref{eq:Mij-integrals} using
the quotient rule

\begin{equation}
\begin{array}{ccc}
\frac{\partial M_{ij}}{\partial A_{k}} & = & \frac{-\lambda_{i}\left[\int f_{j}(E)f_{k}(E)S_{i}(E)e^{-\mathbf{A\cdot f(E)}}dE\right]+\left[\int f_{j}(E)S_{i}(E)e^{-\mathbf{A\cdot f(E)}}dE\right]\left[\int f_{k}(E)S_{i}(E)e^{-\mathbf{A\cdot f(E)}}dE\right]}{\lambda_{i}^{2}}\\
 & = & -\left[\left\langle f_{j}(E)f_{k}(E)\right\rangle _{i}-\left\langle f_{j}(E)\right\rangle _{i}\left\langle f_{k}(E)\right\rangle _{i}\right]\\
 & = & M_{ij}M_{ik}-\left\langle f_{j}(E)f_{k}(E)\right\rangle _{i}.
\end{array}\label{eq:dMijdAk}
\end{equation}
\noindent{}In this equation, $\lambda_{i}$ is given by Eq. \ref{eq:lambda-k}
and $\left\langle \right\rangle _{i}$ is defined in Eq. \ref{eq:M-ij-fj-eff}.

The effective values of $\mathbf{M}$ depend on  the transmitted spectrum
and are, for most spectra, a function of $\mathbf{A}$. If the spectra
are single delta functions at energy $E_{1},\ E_{2}$ such as from
an isotope source or a synchrotron source, then 

\begin{equation}
\left\langle f_{j}(E)f_{k}(E)\right\rangle _{i,delta\ function}=M_{ij}M_{ik}\label{eq:<mimj>delta}
\end{equation}
and $\nicefrac{\partial M_{ij}}{\partial A_{k}}=0$. With single delta
function spectra, there is no beam hardening and the $M_{ij}$ are
constants for all values of $\mathbf{A}$. Notice that Eq. \ref{eq:dMijdAk}
has the same form as a covariance.

\subsection{Analytical formula for endpoints of zero determinant line\label{sub:Analytical-formulas}}

Since $\mathbf{M}$ is a $2\times2$ matrix, its determinant is 
\begin{equation}
\det\left(\mathbf{M}\right)=M_{11}M_{22}-M_{12}M_{21}\label{eq:detM}
\end{equation}
\noindent{}To apply the inversion theorem, we need to determine whether
the determinant is zero in the first quadrant of the $\mathbf{A}$
plane. The simulations in Sec. \ref{sub:Simulations-of-det(M(A))}
show that the product terms $M_{ij}M_{pq}(\mathbf{A})$ are approximately
planar and do not intersect if the transformation is invertible. If
the transformation is not invertible, the $M_{11}M_{22}(\mathbf{A})$
and $M_{12}M_{21}(\mathbf{A})$ surfaces do intersect and the curve
of intersection is approximately a straight line. One approach to
develop analytical formulas for invertibility is to determine the
line by points of zero determinant on the $A_{1}$ and $A_{2}$ axes.
the intersections of the the product terms on the $A_{1}$ and $A_{2}$
axes. If the determinant is zero in the first quadrant so the system
is invertible, both of the intersections will be not physically feasible,
that is, the product terms at the intersection are less than or equal
to zero.

Using a linear approximation to the determinant along the $A_{k},\ k=1,2$
axes 
\begin{equation}
\begin{array}{ccc}
\det\left(\mathbf{M}\right)(A_{k}) & \approx & \det\left(\mathbf{M}\right)(\mathbf{0})+\left\{ \frac{\partial}{\partial A_{k}}\left[M_{11}M_{22}\right](\mathbf{0})-\frac{\partial}{\partial A_{k}}\left[M_{12}M_{21}\right](\mathbf{0})\right\} A_{k}.\end{array}\label{eq:detM-linear-approx}
\end{equation}

\noindent{}Setting the right hand side equal to zero and solving
for $A_{k}$ 
\begin{equation}
A_{k,0}=-\frac{\det\left(\mathbf{M}\right)(\mathbf{0})}{\frac{\partial}{\partial A_{k}}\left[M_{11}M_{22}\right](\mathbf{0})-\frac{\partial}{\partial A_{k}}\left[M_{12}M_{21}\right](\mathbf{0})}.\label{eq:Ak0}
\end{equation}
We can use the product rule to evaluate the derivatives in the denominator
\begin{equation}
\begin{array}{ccc}
\frac{\partial}{\partial A_{k}}\left[M_{ij}M_{pq}\right] & = & M_{ij}\frac{\partial M_{pq}}{\partial A_{k}}+M_{pq}\frac{\partial M_{ij}}{\partial A_{k}}\end{array}\label{eq:dMijMpq-prod-rule}
\end{equation}

\subsection{Analytical condition for invertibility\label{sub:Analytical-condition}}

Using a linear approximation, the value of the diagonal product term,
$\left[M_{11}M_{22}\right]$, at $A_{k,0}$ is
\begin{equation}
\left[M_{11}M_{22}\right]_{lin}(A_{k,0})\approx M_{11}M_{22}\left(\mathbf{0}\right)+\frac{\partial}{\partial A_{1}}\left[M_{11}M_{22}\right](\mathbf{0})A_{k,0}.\label{eq:P-A-linear}
\end{equation}

\noindent{}Note that the diagonal and off diagonal product terms
are equal at the $\det(\mathbf{M}$)=0 points. In Sec. \ref{sub:Simulations-of-det(M(A))},
I show numerically that the linear approximation to $\left[M_{11}M_{22}\right]_{lin}(A_{k,0})$
is positive for the non-invertible spectrum and negative for at least
one of the $A_{k}$ axes for the invertible spectra.

We can derive an analytical formula for $\left[M_{11}M_{22}\right]_{lin}(A_{k,0})$
by substituting $A_{k,0}$ from Eq. \ref{eq:Ak0} into Eq. \ref{eq:P-A-linear}
and simplifying:\begin{Large}
\begin{equation}
\begin{array}{cc}
\left[M_{11}M_{22}\right]_{lin}(A_{k,0})=\ldots\\
\\
\frac{M_{11}M_{22}M_{12}M_{21}\left[\frac{\left\langle f_{2}f_{k}\right\rangle _{1}}{M_{12}}-\frac{\left\langle f_{1}f_{k}\right\rangle _{1}}{M_{11}}+\frac{\left\langle f_{1}f_{k}\right\rangle _{2}}{M_{21}}-\frac{\left\langle f_{2}f_{k}\right\rangle _{2}}{M_{22}}\right]}{\det\left(\mathbf{M}\right)(\mathbf{0})(M_{1k}+M_{2k})+M_{21}\left\langle f_{2}f_{k}\right\rangle _{1}-M_{22}\left\langle f_{1}f_{k}\right\rangle _{1}+M_{12}\left\langle f_{1}f_{k}\right\rangle _{2}-M_{11}\left\langle f_{2}f_{k}\right\rangle _{2}}
\end{array}\label{eq:Prod-at-zero-det}
\end{equation}

\end{Large}

The simulations also show that the denominator of \emph{Eq. \ref{eq:Prod-at-zero-det}}
is positive for the three spectra tested. Since we are only interested
in whether the product value is negative or zero and the simulations
show the denominator is always positive, we need consider the sign
of the numerator of \emph{Eq. \ref{eq:Prod-at-zero-det}}\begin{Large}
\begin{equation}
Numerator=M_{11}M_{22}M_{12}M_{21}\left[\frac{\left\langle f_{2}f_{k}\right\rangle _{1}}{M_{12}}-\frac{\left\langle f_{1}f_{k}\right\rangle _{1}}{M_{11}}+\frac{\left\langle f_{1}f_{k}\right\rangle _{2}}{M_{21}}-\frac{\left\langle f_{2}f_{k}\right\rangle _{2}}{M_{22}}\right].\label{eq:num-diagProd-0}
\end{equation}

\end{Large}

\noindent{}The elements of $\mathbf{M}$ are effective values of
the basis functions and are always positive so the sign is determined
by the expression in brackets of Eq. \ref{eq:num-diagProd-0}.

Notice that for delta function spectra the first and second and the
third and fourth terms of the numerator cancel it is equal to zero.
Similarly, the denominator is also zero for delta functions so the
product term is indeterminate $\frac{0}{0}$ and the system will be
invertible if the Jacobian determinant at zero thickness is not equal
to zero.

\subsection{Simulations of det(M(A))\label{sub:Simulations-of-det(M(A))}}

In the following sections, I will study by computer simulation whether
the Jacobian determinant is zero inside $C$ for several spectra:
(1) a set that are known to be non-invertible for two specific A-vectors,
(2) a similar set that are better conditioned, and (3) x-ray tube
spectra with different voltages.

\subsubsection{3 energy non-invertible}

\begin{figure}[H]
\centering{}(a)\includegraphics[scale=0.6]{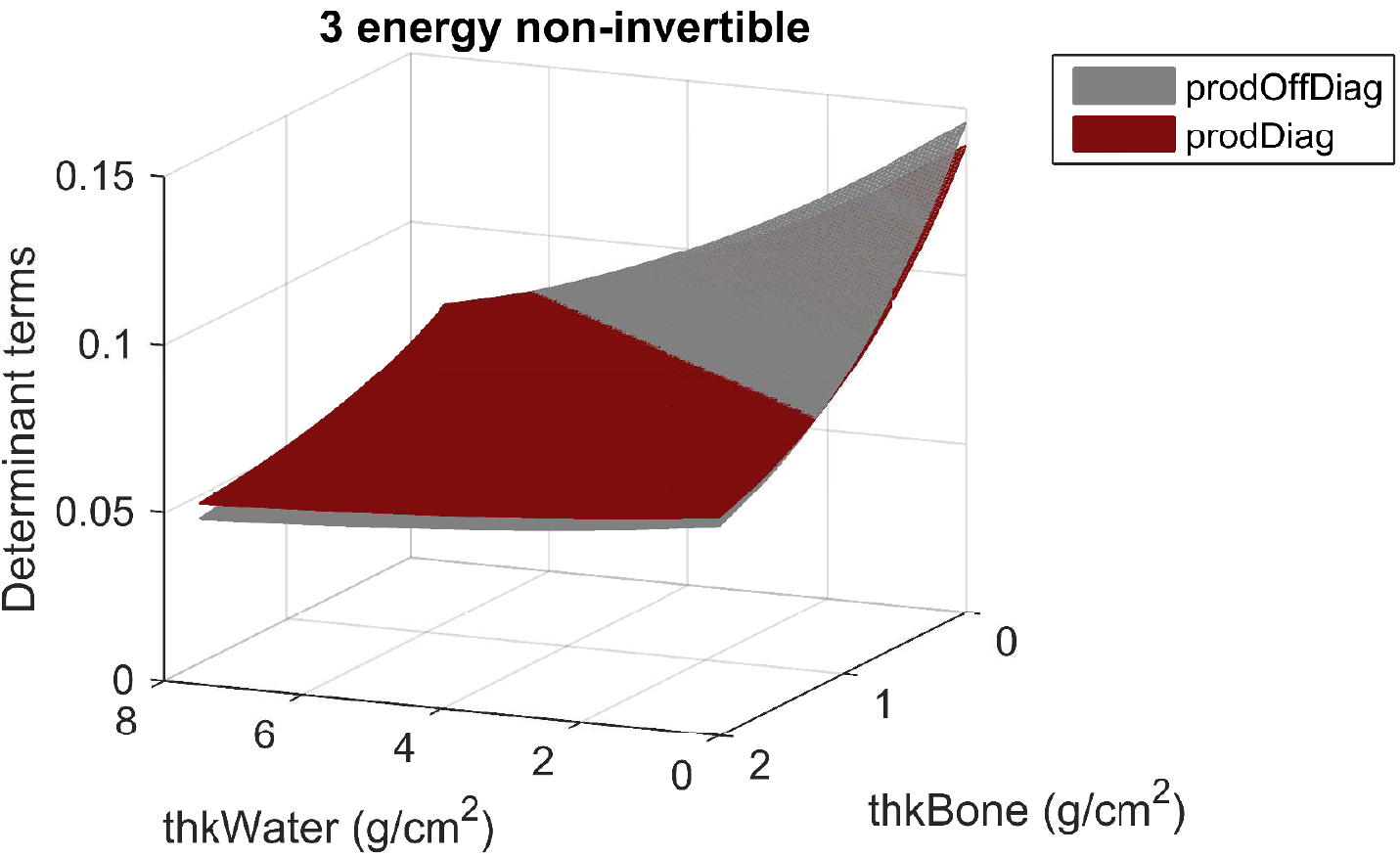}(b)\includegraphics[scale=0.5]{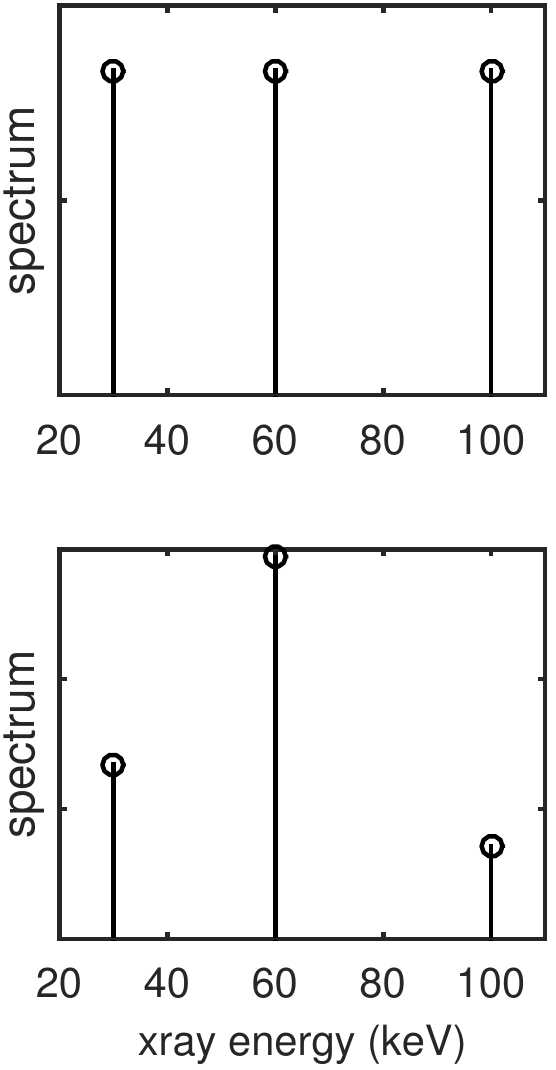}

\protect\caption{Three dimension plot (a) of the products of the diagonal and off-diagonal
elements of the determinant $\det(M\left(\mathbf{A}\right))=M_{11}M_{22}-M_{12}M_{21}$
as a function of $\mathbf{A}$ for the Levine non-invertible three
energy spectrum. The determinant is zero if the surfaces intersect.
Each product term is plotted as a separate color-coded surface as
shown in the legend. Part (b) is a plot of the spectra as a function
of x-ray energy. \label{fig:3D-detM-levine}}
\end{figure}

Levine\cite{levine2017nonuniqueness} described a set of spectra that
he found were not one-to-one. In this section, we analyze these spectra
using the methods described in the previous sections. Fig. \ref{fig:3D-detM-levine}
(a) shows a three dimension plot of the diagonal and off-diagonal
$\mathbf{M}$ elements product terms. Since the determinant is $\det(M)=M_{11}M_{22}-M_{12}M_{21}$,
it is zero if the surfaces intersect. Each term is plotted as a separate
color-coded surface as shown in the legend of the figure. Part (b)
of the figure shows the spectra. From the figure, it is clear that
the product term surfaces do intersect on approximately a straight
line.

\begin{figure}[H]
\centering{}(a)\includegraphics[scale=0.5]{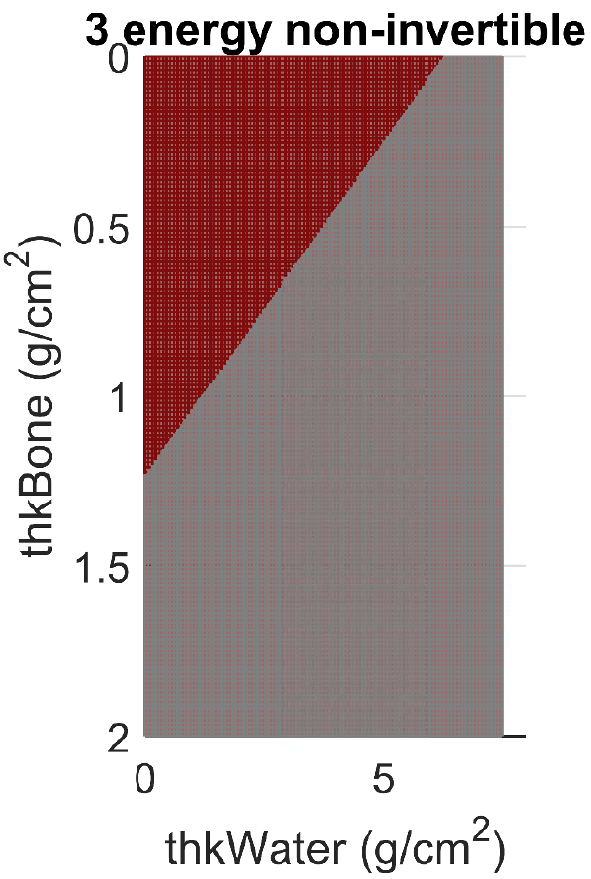} (b)\includegraphics{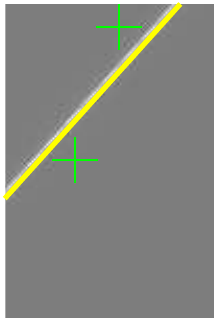}

\protect\caption{Projection of the terms of $det(M(A))$ onto the A-plane. Part (a)
is a view of the diagonal and off-diagonal determinant terms in Fig.
\ref{fig:3D-detM-levine} projected along the Z-axis. Part (b) is
an image where the grayscale is proportional to $\nicefrac{1}{\left|det(M(A))\right|}$.
The bright line in the image shows the zero values. A straight line
was fit to the line of maxima and is plotted on the image in yellow.
The close fit of the straight line shows that the surfaces are close
to planes so their intersection is a line. Also shown is the Levine
ambiguous A-vectors plotted as the green crosses. \label{fig:NonInvertLines}}
\end{figure}

Fig. \ref{fig:NonInvertLines} shows views of the diagonal and off-diagonal
$\mathbf{M}$ elements product data projected onto the A-plane. Part
(a) is a view of the diagonal and off-diagonal determinant terms in
Fig. \ref{fig:3D-detM-levine} projected along the Z-axis. Part (b)
is an image where the grayscale is proportional to $\nicefrac{1}{\left|det(M(A))\right|}$.
The bright line in the image shows the zero values. A straight line
was fit to the line of maxima and is plotted on the image in yellow.
The close fit of the straight line shows that the surfaces are close
to planes so their intersection is a line.

Fig. \ref{fig:NonInvertLines} also shows the ambiguous A-vectors
found by Levine plotted as the green crosses. Notice that the points
do not fall on the line of zero determinant values.

\begin{figure}[H]
\centering{}\includegraphics[scale=0.6]{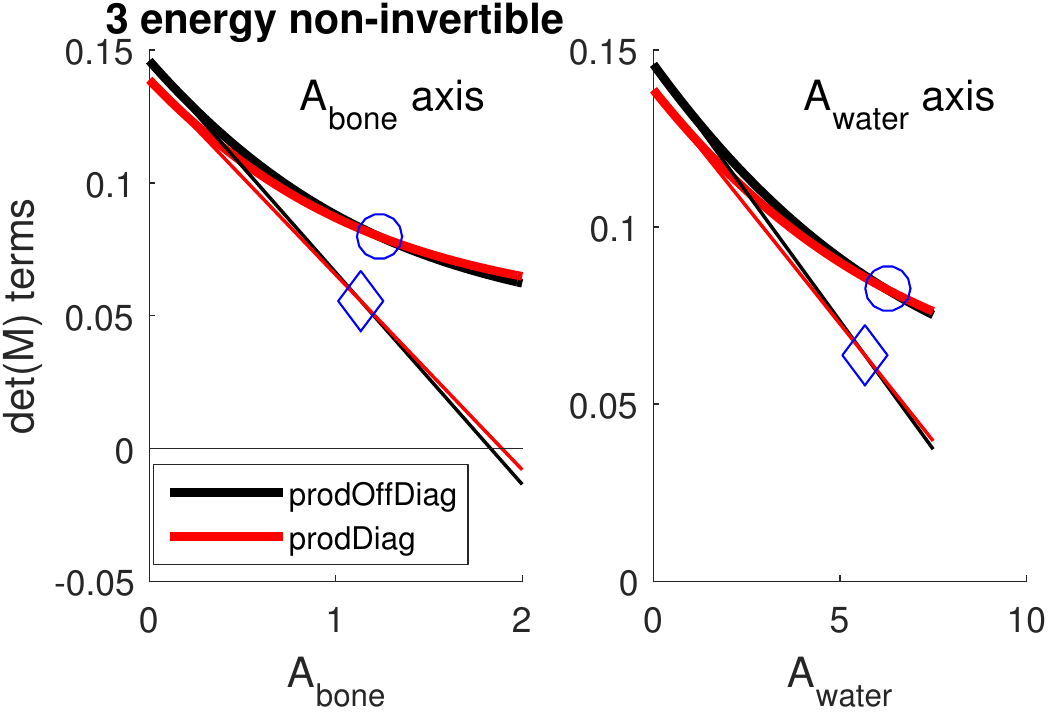}

\protect\caption{Plot of the diagonal and off-diagonal surfaces along the $A_{1}$
and $A_{2}$ axes. The curved lines are the actual surfaces with a
circle at their intersection where $\det(M)$ is zero. Also shown
are straight line fits to the curved lines using the slope at the
origin. The slope was computed with Eq. \ref{eq:dMijMpq-prod-rule}.
The intersection of the fit lines is the diamond shape. \label{fig:Prod-diag-axes}}

\end{figure}

Fig. \ref{fig:Prod-diag-axes} shows one dimensional plots of the
diagonal and off-diagonal determinant terms on the $A_{bone}$ and
$A_{water}$ axes. The curved lines are the actual surfaces with a
circle at their intersection where $\det(M)$ is zero. Also shown
are straight line fits to the curved lines using the slope at the
origin. The slope was computed with Eq. \ref{eq:dMijMpq-prod-rule}.
The intersection of the fit lines is the diamond shape. 

Once the zero determinant points are found on the $A_{k}$ axes, we
can use them to determine the zero determinant region as the line
joining the intersection points. This is the yellow line in part (b)
of Fig. \ref{fig:NonInvertLines}. The yellow line and the region
of large inverse determinant coincide.

\subsubsection{3 energy invertible}

\begin{figure}[H]
\centering{}(a)\includegraphics[scale=0.6]{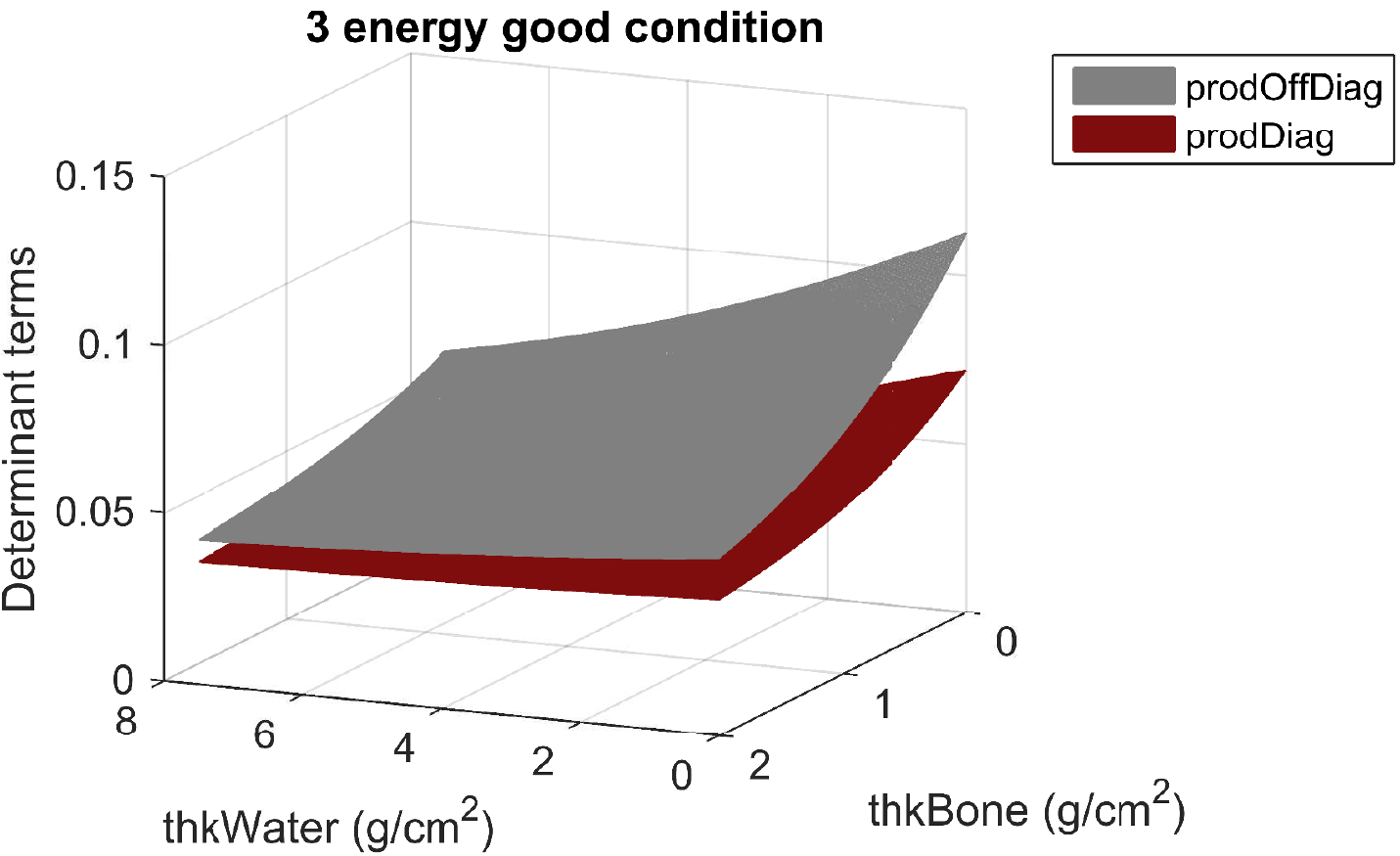}(b)\includegraphics[scale=0.5]{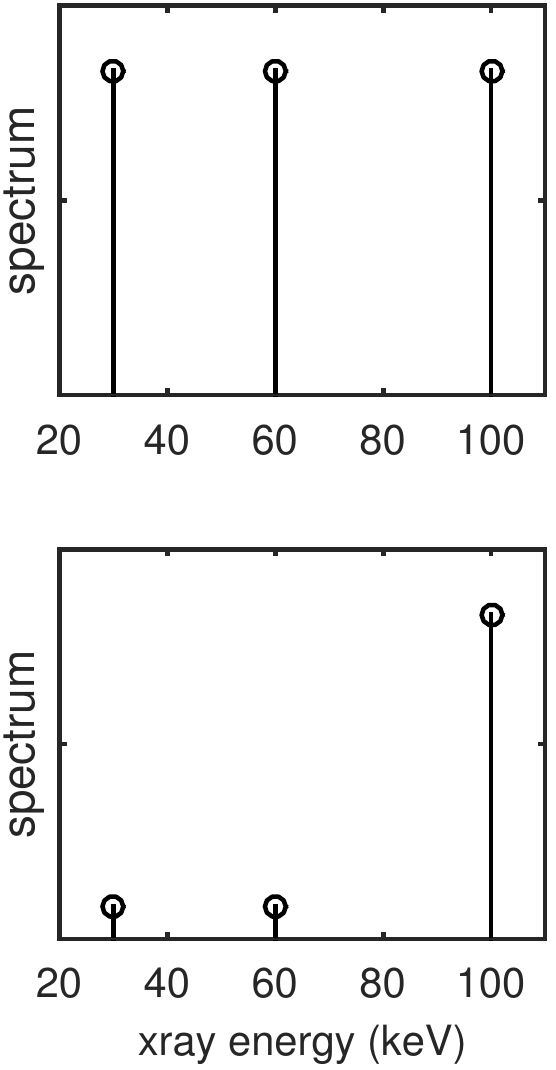}

\protect\caption{Three dimension plot of the products of the diagonal and off-diagonal
elements of $\mathbf{\det(M})$ as a function of $\mathbf{A}$ for
well-conditioned three energy spectra. Each product term is plotted
as a separate color-coded surface as shown in the legend. Notice that
the surfaces do not intersect so the determinant is not equal to zero
in the first quadrant of the A-plane. Part (b) is a plot of the spectra.
\label{fig:3D-det-3spike-Good_cond}}
\end{figure}

Fig. \ref{fig:3D-det-3spike-Good_cond} is a three dimension plot
of the products of the diagonal and off-diagonal elements of $\mathbf{\det(M})$
as a function of $\mathbf{A}$ for well-conditioned three energy spectra.
Part (b) is a plot of the spectra. Each product term is plotted as
a separate color-coded surface as shown in the legend. Notice that
the surfaces do not intersect so the determinant is not equal to zero
in the first quadrant of the A-plane. 

\begin{figure}[H]
\centering{}\includegraphics[scale=0.8]{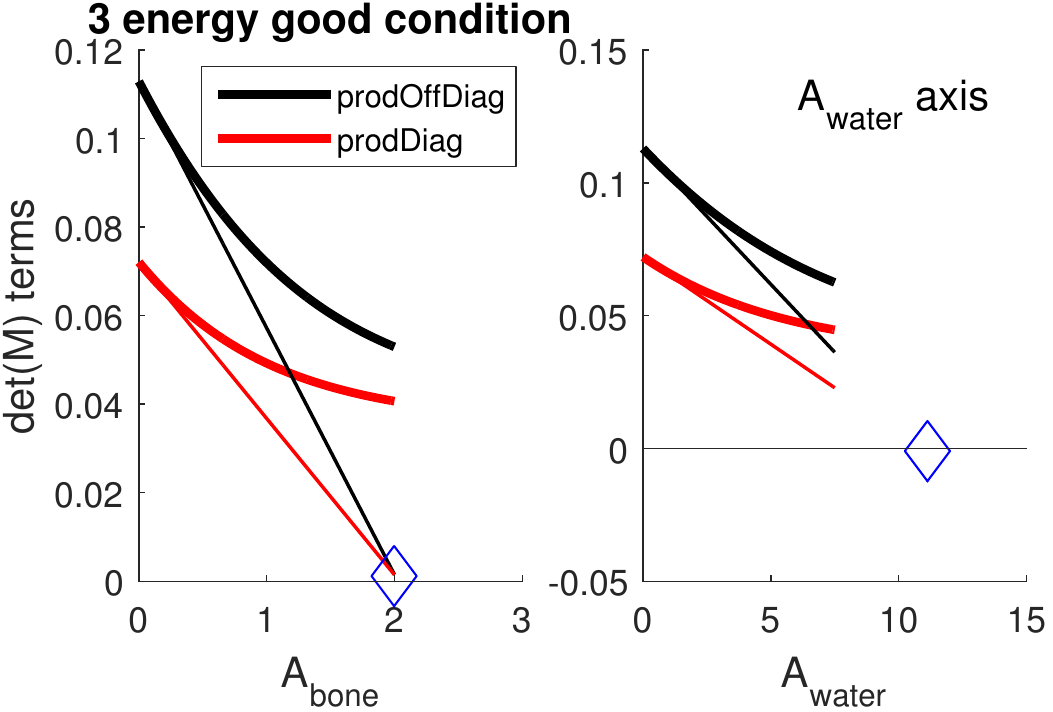}

\protect\caption{Plots of the diagonal and off-diagonal surfaces along the $A_{1}$
and $A_{2}$ axes for invertible 3 energy spectra. Notice that the
actual product terms, shown by the curved lines, do not intersect.
Also shown are the straight lines fit to the derivative at the origin.
Their intersection, shown by the diamond icon, occurs outside the
region of interest and the product term values at the intersection
are equal to zero, which is physically not possible. \label{fig:A-3spike-good}}

\end{figure}

Fig. \ref{fig:A-3spike-good} shows plots of the diagonal and off-diagonal
surfaces along the $A_{1}$ and $A_{2}$ axes for invertible 3 energy
spectra. Notice that the actual product terms, shown by the curved
lines, do not intersect. Also shown are the straight lines fit to
the derivative at the origin. Their intersection, shown by the diamond
icon, occurs outside the region of interest and the product term values
are equal to zero, which is mathematically not possible.

\subsubsection{Voltage switched x-ray tube spectra}

\begin{figure}[H]
\centering{}(a) \includegraphics[scale=0.6]{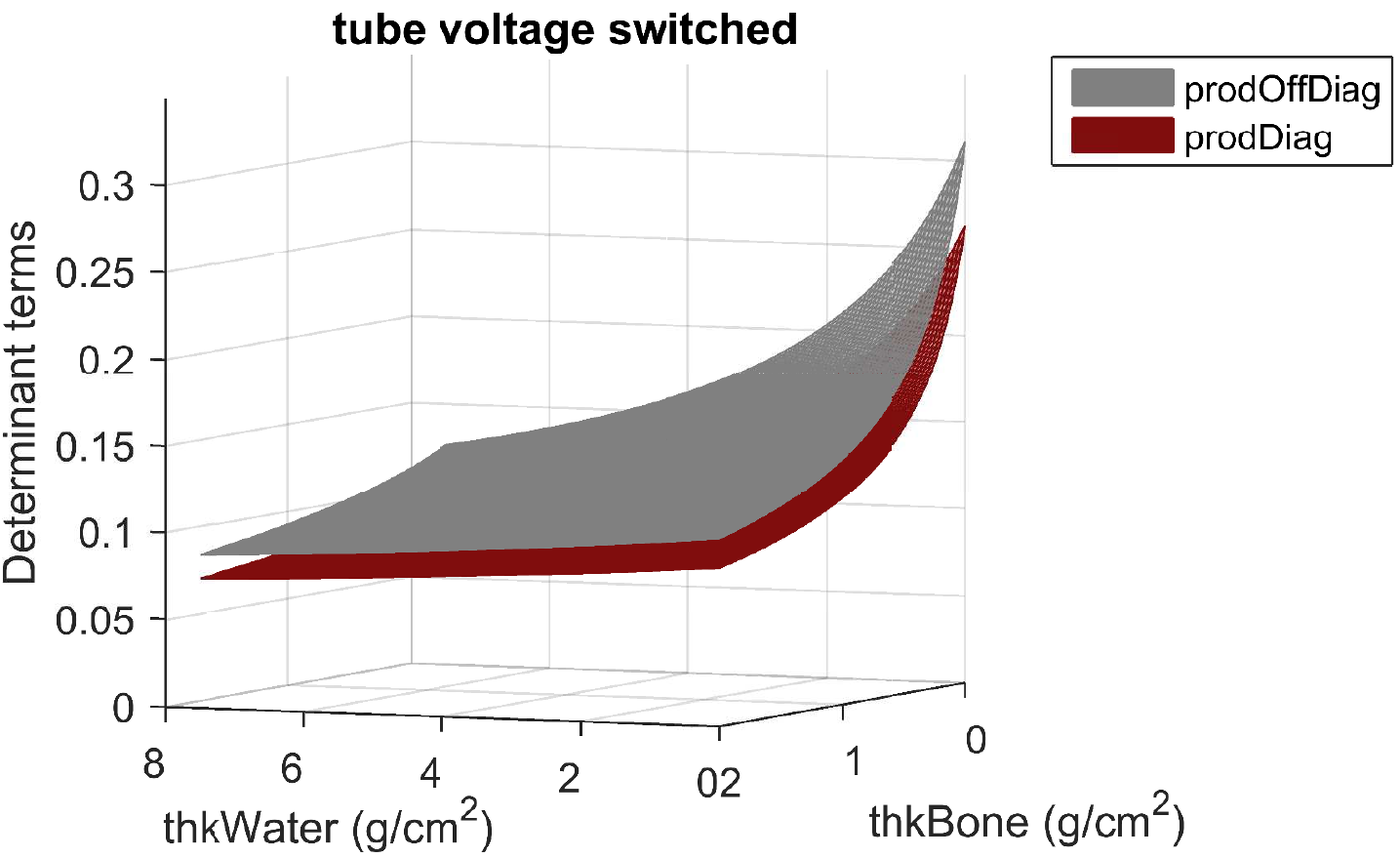} (b) \includegraphics[scale=0.5]{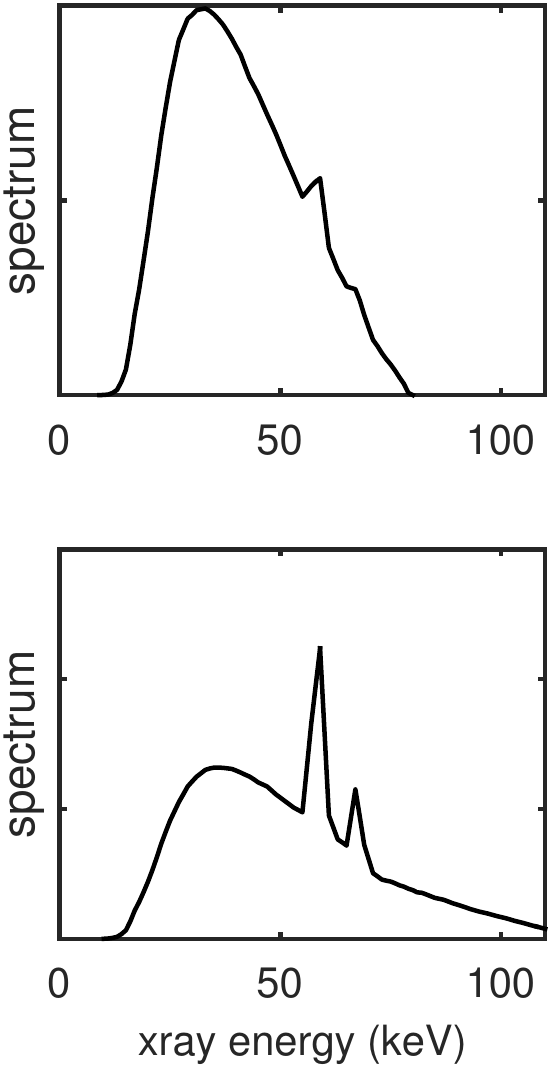}

\protect\caption{Three dimension plot of the products of the diagonal and off-diagonal
elements of $\mathbf{\det(M})$ as a function of $\mathbf{A}$ for
voltage switched (80 and 120 kV) x-ray tube spectra. Notice that the
surfaces do not intersect so the determinant is not equal to zero
in the first quadrant of the A-plane. Part (b) is a plot of the spectra.
\label{fig:3D-voltage-switched}.}

\end{figure}

Fig. \ref{fig:3D-voltage-switched} shows three dimension plot of
the diagonal and off-diagonal product terms of $\mathbf{\det(M})$
for voltage switched (80 and 120 kV) x-ray tube spectra. The surfaces
do not intersect so the determinant is not equal to zero. 

\begin{figure}[H]
\centering{}\includegraphics[scale=0.8]{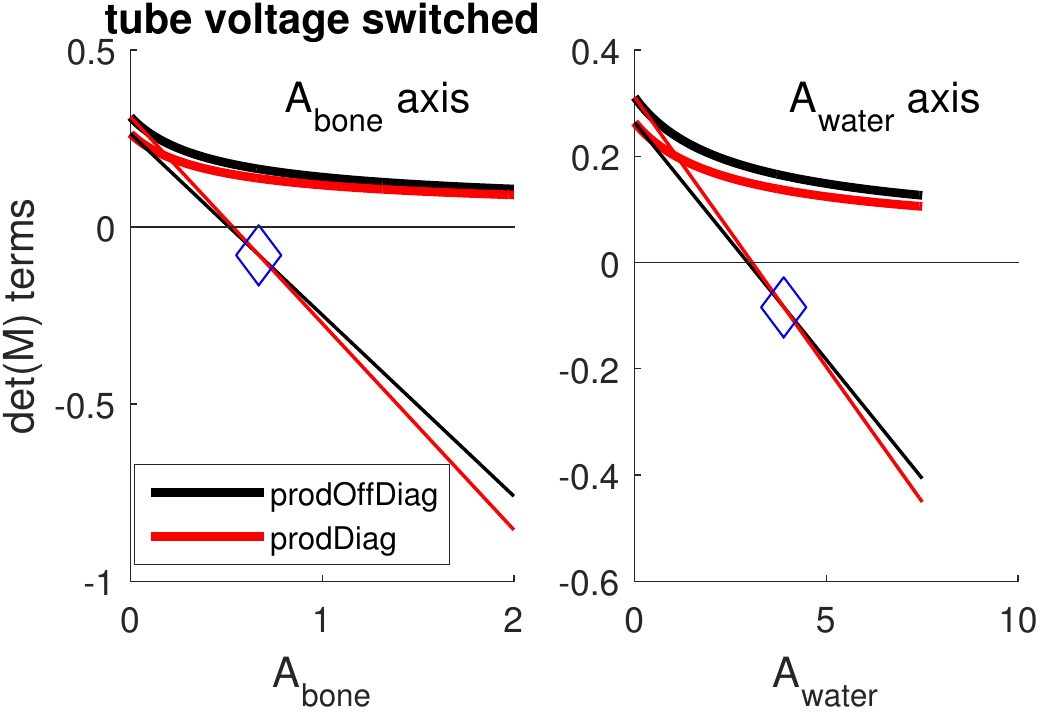}

\protect\caption{Plots of the diagonal and off-diagonal surfaces along the $A_{1}$
and $A_{2}$ axes for the voltage switched spectra. The actual product
terms, shown by the curved lines, do not intersect. Also shown are
the straight lines fit to the derivative at the origin. At their intersection,
shown by the diamond icon, the product term values are negative, which
is not physically possible. \label{fig:A0-tube}}

\end{figure}

Fig. \ref{fig:A0-tube} shows plots of the diagonal and off-diagonal
surfaces along the $A_{1}$ and $A_{2}$ axes for the voltage switched
spectra. The actual product terms, shown by the curved lines, do not
intersect. Also shown are the straight lines fit to the derivative
at the origin. At their intersection, shown by the diamond icon, the
product term values are negative, which is not physically possible.

\subsubsection{$\mathbf{L}$ data for non-invertible spectra}

\begin{figure}[h]
\centering{}(a) \includegraphics[scale=0.6]{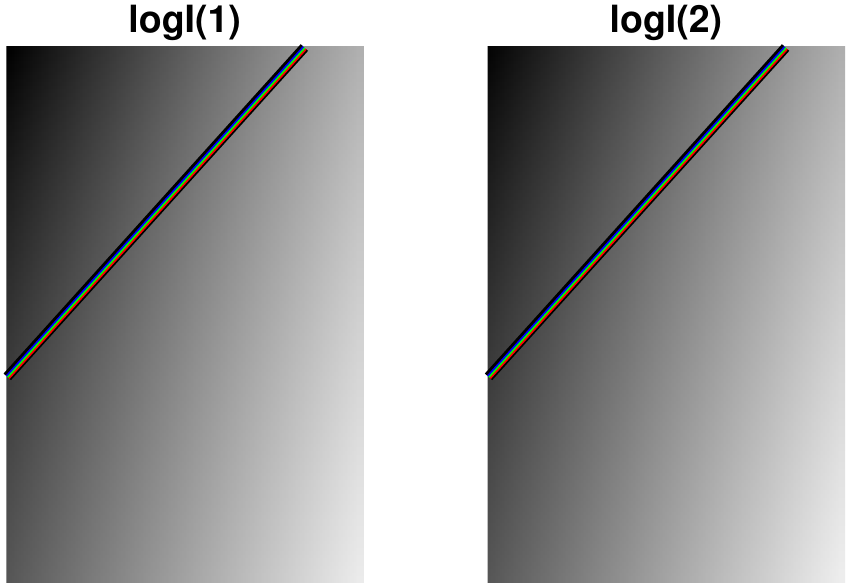} (b) \includegraphics[scale=0.6]{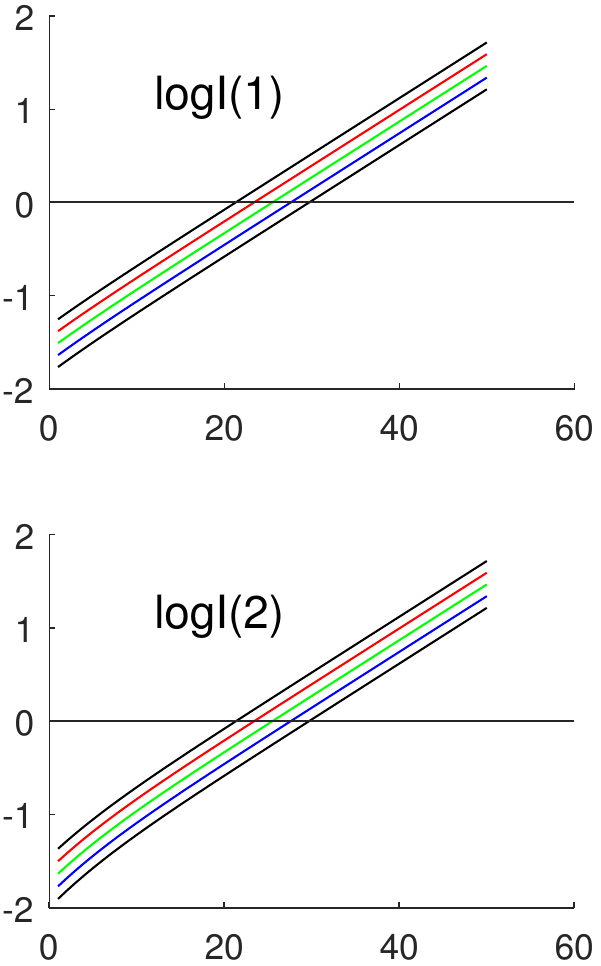}

\protect\caption{$\mathbf{L}$ data for non-invertible 3 energy spectra. Part (a) shows
images where the gray scale is proportional to the $\mathbf{L}$ components.
The images show the line where the determinant is zero and several
lines parallel to it. Part (b) has plots of the $\mathbf{L}$ data
on the lines. The values on the central point of the zero determinant
line are subtracted for ease of display. Notice that the $\mathbf{L}$
data are not constant on the $det(M(A))=0$ line. \label{fig:A-3spike-non-invert}}
\end{figure}

Fig. \ref{fig:A-3spike-non-invert} shows the $\mathbf{L}$ data for
non-invertible 3 energy spectra. Part (a) shows images where the gray
scale is proportional to the $\mathbf{L}$ components. The images
show the line where the determinant is zero and several lines parallel
to it. Part (b) has plots of the $\mathbf{L}$ data on the lines.
The values on the central point of the zero determinant line are subtracted
for ease of display. Notice that the $\mathbf{L}$ data are not constant
on the $det(\mathbf{M(A)})=0$ line.

\subsection{Invert transform with iterative algorithm}

The effect of the zero values of the Jacobian determinant on the invertibility
of $\mathbf{L(A)}$ was tested with an iterative inverse transform
algorithm. The algorithm iteratively found the value of $A$ that
minimized $\left|\mathbf{L(A)-L_{input}}\right|^{2}$. It was implemented
using the Matlab $fminsearch$ function.

\begin{figure}
\centering{}(a) \includegraphics[scale=1.1]{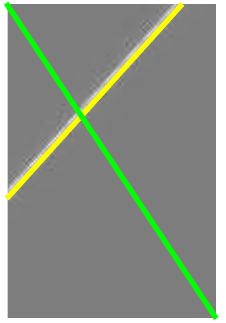}\includegraphics[scale=0.6]{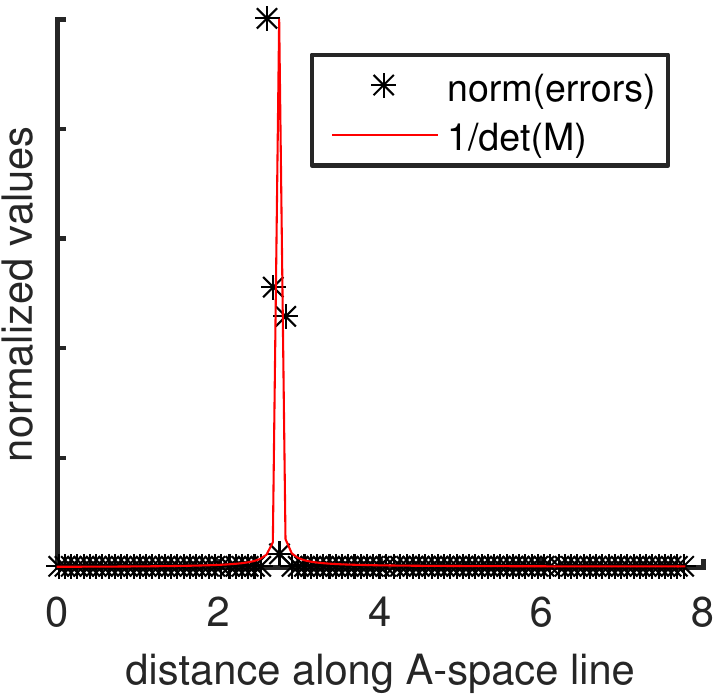}

\protect\caption{Effect of $det(\mathbf{M(A)})$ on errors with iterative inverse transform
algorithm. The norm of the errors of the iterative algorithm for $\mathbf{A}$
values on the green line in panel (a) were computed. They were plotted
as a function of distance along the line in panel (b). The $\det(M)$
was zero on the yellow line in panel (a) and the large errors in (b)
are near the point of intersection of the green and yellow lines.\label{fig:Iter-errors}}

\end{figure}

The norm of the errors of the iterative algorithm for $\mathbf{A}$
values on the green line in Fig. \ref{fig:Iter-errors}(a) were computed.
They were plotted as a function of distance along the line in Fig.
\ref{fig:Iter-errors}(b). The $\det(M)$ was zero on the yellow line
in Fig. \ref{fig:Iter-errors}(a) and the large errors in Fig. \ref{fig:Iter-errors}(b)
are near the point of intersection of the green and yellow lines.

\subsubsection{Iterative algorithm with Levine ambiguous points}

The iterative algorithm was also used to test the ambiguous $\mathbf{A}$
vectors noted by Levine\cite{levine2017nonuniqueness}, $[2.5,\ 1]$
and $[4.09,\ 0.13]$. Using the $\mathbf{L}$ vector computed for
the first $\mathbf{A}$ as input to the iterative algorithm, the solution
depended on the initial point of the iteration. If the point was near
one of the two values, the algorithm converged to that value. If the
point was far from both, for example the origin, the algorithm converged
to $[2.5,\ 1]$.

\subsection{\label{sub:Values-of-det-terms}Values of determinant terms with
invertible and non-invertible spectra}

As discussed in Sec. \ref{sub:Analytical-condition}, the analytical
formula for signs of values of the product terms at the intersections
of the line of zero Jacobian values determines whether the spectra
are invertible. The formula was tested with the simulation data described
in Sec. \ref{sub:Simulations-of-det(M(A))} and the results for the
product terms for the three spectra tested are shown in Table \ref{tab:Determinant-product-terms}. 

\begin{table}[H]
\protect\caption{Determinant product terms at intersections of zero Jacobian line with
$A_{1}$ and $A_{2}$ axes. \label{tab:Determinant-product-terms}}

\centering{}%
\begin{tabular}{|l|c|c|}
\hline 
 & \textbf{$A_{1}$} & \textbf{$A_{2}$}\tabularnewline
\hline 
\textbf{3 Energy Non-invertible} & 0.055661 & 0.063935\tabularnewline
\hline 
\textbf{3 Energy Good Condition} & 0.0013775 & -0.0011822\tabularnewline
\hline 
\textbf{Tube Voltage Switched} & -0.077947 & -0.084209\tabularnewline
\hline 
\end{tabular}
\end{table}

The theoretical results indicated that the sign of the product terms
is determined by the numerator of Eq. \ref{eq:Prod-at-zero-det}.
Table \ref{tab:Numerator-and-denominator} shows the numerator and
denominator of the product terms equation for the tested spectra.

\begin{table}[H]

\protect\caption{Numerator and denominator of Eq. \ref{eq:Prod-at-zero-det} for the
product terms of the tested spectra. \label{tab:Numerator-and-denominator}}

\centering{}%
\begin{tabular}{|l|c|c|}
\hline 
Numerator $M_{11}M_{22}$ & \textbf{$A_{1}$} & \textbf{$A_{2}$}\tabularnewline
\hline 
\textbf{3 Energy Non-invertible} & 0.00035307 & 8.1111e-05\tabularnewline
\hline 
\textbf{3 Energy Good Condition} & 2.8091e-05 & -4.3262e-06\tabularnewline
\hline 
\textbf{Tube Voltage Switched} & -0.0055665 & -0.0010313\tabularnewline
\hline 
\end{tabular}

$\ $

$\ $

\centering{}%
\begin{tabular}{|l|c|c|}
\hline 
Denominator $M_{11}M_{22}$ & \textbf{$A_{1}$} & \textbf{$A_{2}$}\tabularnewline
\hline 
\textbf{3 Energy Non-invertible} & 0.0063432 & 0.0012686\tabularnewline
\hline 
\textbf{3 Energy Good Condition} & 0.020393 & 0.0036596\tabularnewline
\hline 
\textbf{Tube Voltage Switched} & 0.071413 & 0.012247\tabularnewline
\hline 
\end{tabular}
\end{table}

\section{Discussion}

The invertibility of spectral measurements is an important issue.
Fig. \ref{fig:3D-det-3spike-Good_cond} shows that the errors with
the iterative algorithm show a sharp increase near the line of non-invertibility.
My previous results with a three dimension system \cite{alvarez2017noninvertibility}
showed that the noise increased by a huge factor near the plane of
non-invertibility. X-ray system designers need tools to determine
whether their system may become non-invertible in some part of the
A-vector operating range because the system may become unusable near
the region of non-invertibility. 

In the results shown here, the non-invertible region is nearly a straight
line in the two dimension A-space. In my previous paper\cite{alvarez2017noninvertibility}
the non-invertible region for a three dimension basis set is a plane.
The non-invertibility regions for both cases may be explained by the
observation that the elements of the $\mathbf{M}$ matrix are nearly
linear as a function of $\mathbf{A}$. The results in Figures \ref{fig:3D-detM-levine},
\ref{fig:3D-detM-levine}, and \ref{fig:3D-det-3spike-Good_cond}
show that the terms in the determinant of $\mathbf{M}$ are also close
to linear so their intersection, which results in a zero determinant
value, is close to a straight line in two dimensions or a plane in
three dimensions. 

Sections \ref{sub:Simulations-of-det(M(A))} and \ref{sub:Analytical-condition}
derive analytical formulas for the line of non-invertibility in the
two dimension A-space using a linear approximation to the terms in
the determinant of $\mathbf{M}$. The line is determined by its intersection
with the A axes. For the non-invertible Levine spectrum, Fig.  \ref{sub:Simulations-of-det(M(A))}
shows that the actual and linear approximation intersection points
are physically feasible. That is, at positive values and with positive
values for the determinant terms at the intersection points. For the
invertible three spike spectra, Fig. \ref{sub:Simulations-of-det(M(A))}
shows that the actual determinant terms do not intersect. The linear
approximation does intersect but at the intersection points the determinant
product terms are negative which is not physically feasible. For the
two voltage x-ray tube spectra, Fig. \ref{sub:Simulations-of-det(M(A))}
shows that the actual determinant terms again do not intersect and,
with the linear approximation, the determinant terms are again negative
at the points of intersection.

Table \ref{tab:Determinant-product-terms} gives the values for the
product terms at the intersections with the axes for the three spectra
tested. As indicated by the theory, they are positive for the non-invertible
spectra and at least one is negative for the three energy good condition
and voltage switched spectra, which are both invertible. Table \ref{tab:Numerator-and-denominator}
shows that the denominator of Eq. \ref{eq:Prod-at-zero-det} is always
positive so the sign of the result for the product term is determined
by the numerator and in particular by the expression in brackets in
Eq. \ref{eq:num-diagProd-0}.

Therefore, for the cases tested, the linear approximation gives useful
results and the formulas give insight into the conditions leading
to non-invertibility. There may, of course, be cases where the linear
approximation does not give accurate results. The approach taken was
a trade-off between accuracy and simplicity of the formulas. Simple
formulas allow us to gain physical insight into the invertibility
conditions at the expense of lower accuracy. 

Levine\cite{levine2017nonuniqueness} introduces an alternate approach
to the dual energy inversion problem to that described here. The formulation
focuses on the exponent of the transmission equation, Eq. \ref{eq:Ik-integral}.
It puts the logarithm of the spectral density in the exponent and
combines them with $\mathbf{A\cdot f(E)}$. The combination is used
to derive A-space surfaces specified by the measurement spectra. The
$\mathbf{A}$ vectors that invert the equations are at the intersection
of the surfaces. With a particular set of spectra, Levine found two
A-vector points where the surfaces intersect. The transformation is
therefore not one-to-one and is not invertible. Interestingly, the
two ambiguous points are not on the line of zero determinant. 

The results in Fig. \ref{fig:A-3spike-non-invert} show that the $\mathbf{L}$
data on the line of non-invertibility are not constant. Instead, the
$\mathbf{L}$ data vary but the two product terms of the determinant
balance out resulting in a zero determinant value on this line. 

Research is on-going into more accurate theoretical equations for
invertibility and ways to apply the formulation described in this
paper to clinical systems design.

\section{Conclusion}

An approach to studying the invertibility of spectral measurements
is presented and applied to invertible and non-invertible spectra.
Non-invertible dual energy spectra have zero Jacobian determinant
in the first quadrant of the $\mathbf{A}$ vector plane. For the non-invertible
spectra tested, the Jacobian determinant is equal to zero on a curve
in the $\mathbf{A}$ vector plane that is close to a straight line.
An analytical formula is derived for a linear approximation to the
points where the line crosses the A-axes. If the spectra are invertible,
the intersection points are not physically feasible while for the
non-invertible spectra they are feasible indicating zero Jacobian
determinant values in the first quadrant.


\end{document}